\documentclass[a4paper,onecolumn,accepted=2017-09-19]{quantumarticle}

\usepackage[utf8]{inputenc}
\usepackage[english]{babel}
\usepackage[T1]{fontenc}
\usepackage{amsmath}
\usepackage{bbold}
\usepackage{amsfonts,bm,graphicx,verbatim,mathrsfs,units}
\usepackage{hyperref}
\usepackage{tikz}
\usepackage{lipsum}

\usepackage[numbers,sort&compress]{natbib}

\begin{document}

\title{Quasi-probability distributions for observables in dynamic systems}
\date{\today}
\author{Patrick P. Hofer}
\affiliation{Department of Applied Physics, University of Geneva, 1211 Geneva 4, Switzerland}
\orcid{0000-0001-6036-7291}
\email{patrick.hofer@unige.ch}

\begin{abstract}
We develop a general framework to investigate fluctuations of non-commuting observables. To this end, we consider the Keldysh quasi-probability distribution (KQPD). This distribution provides a measurement-independent description of the observables of interest and their time-evolution. Nevertheless, positive probability distributions for measurement outcomes can be obtained from the KQPD by taking into account the effect of measurement back-action and imprecision. Negativity in the KQPD can be linked to an interference effect and acts as an indicator for non-classical behavior. Notable examples of the KQPD are the Wigner function and the full counting statistics, both of which have been used extensively to describe systems in the absence as well as in the presence of a measurement apparatus. Here we discuss the KQPD and its moments in detail and connect it to various time-dependent problems including weak values, fluctuating work, and Leggett-Garg inequalities. Our results are illustrated using the simple example of two subsequent, non-commuting spin measurements.
\end{abstract}
\onecolumn
\maketitle

\section{Introduction}
\label{sec:introduction}

Quasi-probability distributions such as the Wigner function have been an important tool in quantum mechanics since the early days of the theory \cite{wigner:1932}. Indeed, the Wigner function lies at the heart of the phase-space representation of quantum mechanics which is tightly connected to classical statistical mechanics and provides an alternative route to quantization \cite{qm_phase:book}. In classical statistical mechanics, averages of observables can be described as averages taken with respect to some underlying probability distribution. This probability distribution fully describes the system under consideration and reflects the lack of knowledge of the observer. If one has perfect knowledge of the behavior of all degrees of freedom, there would be no need for probability distributions and measurement outcomes could be predicted in a deterministic manner. {This is not possible in quantum mechanics due to the probabilistic nature of the theory.} Similarly to classical statistical mechanics, the fluctuations in quantum mechanics can be encoded in \textit{quasi}-probability distributions. These distributions can take on negative values which is ultimately a consequence of the non-commutativity of operators, reflecting the fact that a measurement inevitably disturbs all subsequent measurements of observables which do not commute with the first one. Negative quasi-probability distributions therefore reflect the impossibility of accessing observables without disturbing them, indicating a non-classical behavior of the system which requires to take into account the action of the measurement apparatus to predict its outputs.
The concept of negative quasi-probability distributions, or \textit{negativity} in short, is therefore closely related to the concept of contextuality which deals with the dependence of measurement outcomes on the experimental setup \cite{spekkens:2008,ferrie:2008,ferrie:2009,ferrie:2011}.

So far, most quasi-probability distributions describe the quantum state of a system. This holds true for the Wigner function and its generalizations to finite dimensional Hilbert spaces \cite{wootters:1987,gibbons:2004,ferrie:2011}, as well as a number of other distributions used in quantum optics \cite{cahill:1969,walls:book}. Negativity in these distributions indicates non-classicality of the quantum state. However, a state that exhibits negativity in some distribution might not do so in another one \cite{lutkenhaus:1995}, implying that non-classicality is not necessarily a general feature but depends on the way a state is interrogated (see also Refs.~\cite{revzen:2005,spekkens:2008,ferrie:2008,ferrie:2009}). A well studied quasi-probability distribution which does not describe the quantum state of a system but rather its dynamics is the full counting statistics (FCS) \cite{levitov:1996,nazarov:book2,nazarov:2003,bednorz:2010,clerk:2011,bednorz:2012,hofer:2016}. Negativity in the FCS has recently been connected to a peculiar interference effect and it was shown that systems which do not show any negativity in their Wigner function can still show negativity in the FCS \cite{hofer:2016}. This further stresses the fact that systems might exhibit completely classical behavior in some aspects, e.g. their instantaneous states, while exhibiting non-classicality in others, e.g. their dynamics which determine the temporal fluctuations within the system. It would therefore be highly desirable to have access to a quasi-probability distribution which encodes only the aspects of interest of a system.
This is in marked contrast to Refs.~\cite{spekkens:2008,ferrie:2008,ferrie:2009,zhu:2016} which consider a full description of states and measurements in terms of quasi-probabilities.

The goal of the present paper is to introduce a general quasi-probability distribution that can be tailored to the problem at hand and thus provides a general framework for investigating fluctuations of non-commuting observables. Because this distribution is based on the Keldysh formalism, we call it the Keldysh quasi-probability distribution (KQPD). When the observables of interest are the position and momentum operators, the KQPD reduces to the Wigner function. Similarly, when one is interested in the time-integral of an observable, the FCS of the corresponding operator is obtained. Furthermore, the outcome of von Neumann type measurements \cite{vonneumann:book} can be predicted by convolving the corresponding KQPD with the Wigner function of the detectors [cf.~Sec.~\ref{sec:vonneumann}]. This convolution takes into account the imprecision as well as the back-action of the performed measurement. An example is provided by the Husimi $Q$-function which is obtained by convolving the Wigner function with Gaussians and predicts the outcomes of simultaneous but imprecise measurements of position and momentum \cite{stenholm:1992}.

The rest of the article is structured as follows. In Sec.~\ref{sec:keldysh}, we introduce the KQPD, discuss the Wigner function, the FCS, and two subsequent spin measurements as examples, and discuss negativity of the KQPD  as an indicator for non-classical behavior. In Sec.~\ref{sec:vonneumann}, we provide a more physical motivation for investigating the KQPD by connecting it to measurement outcomes which requires the introduction of measurement imprecision and back-action. Before concluding in Sec.~\ref{sec:conclusions}, we discuss applications of the KQPD in Sec.~\ref{sec:applications}, including the distribution of thermodynamic work, (anomalous) weak values, and Leggett-Garg inequalities. Throughout this paper, we set $\hbar=1$.

\section{The Keldysh quasi-probability distribution}
\label{sec:keldysh}
In this section we briefly introduce the Keldysh formalism, a powerful framework for tackling out-of-equilibrium problems, where the KQPD arises quite naturally through its moment generating function. Here we only touch upon the ideas in the Keldysh formalism which are relevant for our discussion. For a detailed introduction to the subject we refer the reader to Ref.~\cite{kamenev:book}. We will start by considering the somewhat trivial example of the statistics of a single observable before we generalize the results to the highly non-trivial scenario of multiple, non-commuting observables at different times.

\subsection{Single observable}

\begin{figure}[t]
\centering
\includegraphics[width=\textwidth]{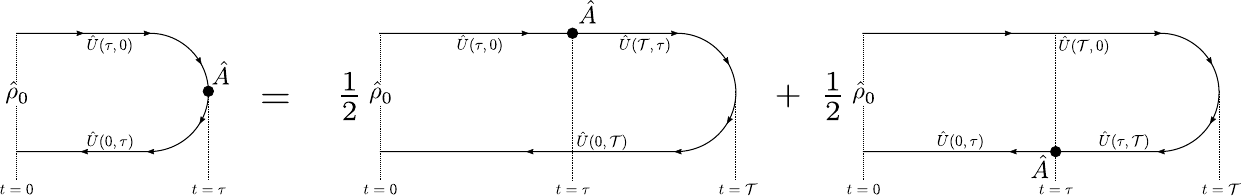}
\caption{Illustration of the closed time contour. The average $\langle\hat{A}\rangle_\tau$ can be obtained by evolving the state (or a mixture of states) up to time $\tau$, where the operator $\hat{A}$ is evaluated, and back again. Alternatively, one can extend this closed-time contour to the Keldysh contour (from $t=0$ to $t= \mathcal{T}$ and back again) and evaluate the operator either on the forward-in-time or on the backward-in-time branch.}
  \label{fig:ctc}
\end{figure}

The main idea that we will be concerned with is the evolution along the closed time contour illustrated in Fig.~\ref{fig:ctc}, an idea that goes back to Schwinger \cite{schwinger:1961}. In this section, we are interested in the statistics of a single observable $\hat{A}$ at time $\tau$. Its average value can be written as
\begin{equation}
\label{eq:avga}
\begin{aligned}
\langle A \rangle_\tau &= {\rm Tr}\left\{\hat{U}(0,\tau)\hat{A}\hat{U}(\tau,0)\hat{\rho}_0\right\} \\&=\frac{1}{2}{\rm Tr}\left\{\hat{U}(0,\mathcal{T})\hat{U}(\mathcal{T},\tau)\hat{A}\hat{U}(\tau,0)\hat{\rho}_0\right\} +\frac{1}{2}{\rm Tr}\left\{\hat{U}(0,\tau)\hat{A}\hat{U}(\tau,\mathcal{T})\hat{U}(\mathcal{T},0)\hat{\rho}_0\right\},
\end{aligned}
\end{equation}
where $\hat{U}(t',t)=\hat{U}^\dag(t,t')$ is the time-evolution operator from time $t$ to time $t'$ and $\hat{\rho}_0$ denotes the density matrix at time $t=0$. Throughout this article, we consider unitary time evolution. Generalizing the results to non-unitary time evolution is in principle straightforward and discussed in App.~\ref{app:nonunitary}. Writing the density matrix as a mixture of pure states, the first line of the last expression can be interpreted as time evolving the states up to time $\tau$, evaluating the operator $\hat{A}$, and evolving the states backward to the initial time $t=0$ [cf.~the left-hand side of Fig.~\ref{fig:ctc}]. On the second line, we inserted identities $\mathbb{1}=\hat{U}(\tau,\mathcal{T})\hat{U}(\mathcal{T},\tau)$, extending the contour along which the state is time evolved. We call the resulting contour, from $t=0$ to $t=\mathcal{T}$ and back again, the Keldysh contour. We can evaluate the operator $\hat{A}$ either along the forward-in-time or along the backward-in-time branch of the contour. As discussed below, the most natural choice is to treat both contours on an equal footing as illustrated graphically in Fig.~\ref{fig:ctc}.

More information on the observable $\hat{A}$ can be obtained by considering its moment generating function given by
\begin{equation}
\label{eq:momgena1}
\Lambda(\lambda;\tau)=\langle e^{-i\lambda\hat{A}}\rangle_\tau = {\rm Tr}\left\{\hat{U}(0,\tau)e^{-i\frac{\lambda}{2}\hat{A}}\hat{U}(\tau,\mathcal{T})\hat{U}(\mathcal{T},\tau)e^{-i\frac{\lambda}{2}\hat{A}}\hat{U}(\tau,0)\hat{\rho}_0\right\}.
\end{equation}
All higher moments can be obtained from this generating function by taking derivatives with respect to $\lambda$
\begin{equation}
\label{eq:momentsa1}
\langle \hat{A}^k\rangle_\tau=\left.i^k\partial_\lambda^k\Lambda(\lambda;\tau)\right|_{\lambda=0}.
\end{equation}
In anticipation of the discussion below, we wrote Eq.~\eqref{eq:momgena1} symmetrically in terms of the Keldysh contour. Taking the derivative with respect to $\lambda$ at $\lambda=0$, we recover Eq.~\eqref{eq:avga}. Since the moment generating function is the average of the exponential of $\hat{A}$, it corresponds to the Fourier transform of a probability distribution associated to the fluctuations of $\hat{A}$
\begin{equation}
\label{eq:probaa1}
\Lambda(\lambda;\tau)=\int_{-\infty}^\infty dA \mathcal{P}(A;\tau)e^{-i\lambda A},\hspace{1cm}\mathcal{P}(A;\tau)=\sum_{A'}\delta(A-A')\langle A'|\hat{U}(\tau,0)\hat{\rho}_0\hat{U}^\dag(\tau,0)|A'\rangle,
\end{equation}
where the sum over $A'$ goes over all eigenstates $\hat{A}|A'\rangle=A'|A'\rangle$ and has to be replaced by an integral if $\hat{A}$ has a continuous spectrum. In the case of a single observable, the KQPD $\mathcal{P}(A;\tau)$ is a positive probability distribution which predicts the outcome of projectively measuring $\hat{A}$ at time $\tau$.

The symmetric treatment with respect to the forward-in-time and the backward-in-time branches of the Keldysh contour might seem like an unnecessary complication at this point. Below, we show how such a treatment for multiple observables leads to a time-ordering prescription which is relevant for von Neumann type measurements. Before discussing multiple observables in detail, we express the moment generating function in terms of the classical and quantum fields obtained by a Keldysh rotation [cf.~Eq.~\eqref{eq:keldyshrot}]. To this end, we assume that the observable $\hat{A}$ has a discrete spectrum and that its eigenstates span the whole Hilbert space, i.e. $\mathbb{1}=\sum_{A}|A\rangle\langle A|.$
We then divide the time evolution along the Keldysh contour in Eq.~\eqref{eq:momgena1} into $2N$ steps ($N$ for the forward-, $N$ for the backward-in-time branch) of infinitesimal length $\delta t$ and insert identities in between the steps. Writing the trace in terms of the eigenstates of $\hat{A}$, we then find
\begin{equation}
\label{eq:momgensuma1}
\Lambda(\lambda;\tau)=\sum_{\boldsymbol{A}_+,\boldsymbol{A}_-}\delta_{A_\mathcal{T}^+,A_\mathcal{T}^-}e^{-i\frac{\lambda}{2}(A_\tau^++A_\tau^-)}\mathcal{A}(\boldsymbol{A}_+)\mathcal{A}^*(\boldsymbol{A}_-)\langle A_0^+|\hat{\rho}|A_0^-\rangle.
\end{equation}
Here, $A_t^{+(-)}$ denotes the state that was inserted at time $t$ on the forward-in-time (backward-in-time) branch, the vectors $\boldsymbol{A}_\pm$ gather all $A_t^\pm$ and the amplitudes along the branches read
\begin{equation}
\label{eq:amplitudesa1}
\mathcal{A}(\boldsymbol{A}_\pm)=\langle A_\mathcal{T}^\pm|\hat{U}(\mathcal{T},\mathcal{T}-\delta t)|A_{\mathcal{T}-\delta t}^\pm\rangle\langle A_{\mathcal{T}-\delta t}^\pm|\cdots |A^\pm_{\delta t}\rangle\langle A_{\delta t}^\pm|\hat{U}(\delta t, 0)|A_0^\pm\rangle.
\end{equation}
The Kronecker delta in Eq.~\eqref{eq:momgensuma1} is due to the fact that the point $t=\mathcal{T}$ connects the two branches (cf.~Fig.~\ref{fig:ctc}). We can write the moment generating function in the form of a path integral
\begin{equation}
\label{eq:pathintegral}
\Lambda(\lambda;\tau)=\int \mathcal{D}[A_t^+,A_t^-]e^{-i\frac{\lambda}{2}(A_\tau^++A_\tau^-)}e^{iS[A_t^+,A_t^-]},
\end{equation}
where the action $S$ is determined by $\exp(iS[A_t^+,A_t^-])=\mathcal{A}(\boldsymbol{A}_+)\mathcal{A}^*(\boldsymbol{A}_-)\langle A_0^+|\hat{\rho}|A_0^-\rangle$
and the integration measure reads
$\int \mathcal{D}[A_t^+,A_t^-]=\sum_{\boldsymbol{A}_+,\boldsymbol{A}_-}\delta_{A_\mathcal{T}^+,A_\mathcal{T}^-}$.
Finally, we perform a Keldysh rotation by introducing the new fields \cite{kamenev:book}
\begin{equation}
\label{eq:keldyshrot}
A^{cl}_t=\frac{1}{2}(A_t^++A_t^-),\hspace{2cm} A^q_t = A_t^+-A_t^-,
\end{equation}
where the superscript stands for classical and quantum respectively. This nomenclature derives from the fact that the forward- and backward-in-time evolutions are equal for classical processes. Using these new integration variables, the path integral in Eq.~\eqref{eq:pathintegral} reduces to
\begin{equation}
\label{eq:pathintegral2}
\Lambda(\lambda;\tau)=\int \mathcal{D}[A_t^{cl},A_t^q]e^{-i\lambda A_\tau^{cl}}e^{iS[A_t^{cl},A_t^q]}.
\end{equation}
The corresponding probability distribution then reads
\begin{equation}
\label{eq:probapathint}
\mathcal{P}(A;\tau)=\frac{1}{2\pi}\int_{-\infty}^{\infty}e^{i\lambda A}d\lambda\Lambda(\lambda;\tau)=\int \mathcal{D}[A_t^{cl},A_t^q]\delta(A-A_\tau^{cl})e^{iS[A_t^{cl},A_t^q]}.
\end{equation}
The moment generating function thus describes the fluctuations of $A_\tau^{cl}$ and the probability density $\mathcal{P}(A;\tau)$ is given by the sum over all paths for which $A_\tau^{cl}=A$, weighted by the complex exponential of the action in complete analogy to Feynman path integrals \cite{altland:book}. In the case of a single observable, the quantum field is zero (i.e. $A_\tau^+=A_\tau^-$), making its introduction seem like an unnecessary complication. However, when multiple observables are involved, the quantum fields are no longer bound to be zero. Considering the fluctuations of the classical fields then corresponds to a given time-ordering prescription. We now generalize the results of this section to an arbitrary number of (possibly non-commuting) observables at arbitrary times.

\subsection{Multiple observables}

The KQPD, denoted by $\mathcal{P}(\boldsymbol{A})$, can be defined via its moment generating function which is obtained by generalizing Eq.~\eqref{eq:momgena1} and reads
\begin{equation}
\label{eq:kmomgen}
\begin{aligned}
\Lambda(\boldsymbol{\lambda})&\equiv\int d\boldsymbol{\lambda}e^{-i\boldsymbol{\lambda}\cdot\boldsymbol{A}}\mathcal{P}(\boldsymbol{A})\\
&\equiv{\rm Tr}\left\{\hat{\mathcal{T}}_+\exp\left[-\frac{i}{2}\int_0^\mathcal{T} d\tau\sum_lf_l(\tau)\lambda_l\hat{A}_l(\tau)\right]\hat{\rho}_0\hat{\mathcal{T}}_-\exp\left[-\frac{i}{2}\int_0^\mathcal{T} d\tau\sum_lf_l(\tau)\lambda_l\hat{A}_l(\tau)\right]\right\}\\&
={\rm Tr}\left\{\hat{\mathcal{T}}_K\exp\left[-\frac{i}{2}\int_K d\tau\sum_lf_l(\tau)\lambda_l\hat{A}_l(\tau)\right]\hat{\rho}_0\right\}.
\end{aligned}
\end{equation}
Here $\hat{\mathcal{T}}_{+(-)}$ is the (anti) time-ordering operator, the variables $\lambda_l$ and $A_l$ are grouped in vectors $\boldsymbol{\lambda}$ and  $\boldsymbol{A}$, and the operators are given in the Heisenberg picture, i.e.
\begin{equation}
\label{eq:aheisen}
\hat{A}_l(t)=\hat{U}(0,t)\hat{A}_l\hat{U}(t,0),
\end{equation}
where $\hat{A}$ denotes the operator in the Schrödinger picture (where all considered operators are time-independent). In the last line of Eq.~\eqref{eq:kmomgen}, $\hat{\mathcal{T}}_K$ denotes time-ordering along the Keldysh contour (from $t=0$ to $t=\mathcal{T}$ and back again). The functions $f_l(\tau)$ take into account a possibly time-dependent coupling to the observables $\hat{A}_l$. When using the Keldysh contour, the functions $f_l$ take on the same values on the forward- and the backward-in-time branch.
The case of $N$ observables probed instantaneously at different times $\tau_l$ is obtained by taking $f_l(\tau_l)=\delta(t-\tau_l)$ (for $l\leq N$) resulting in the moment generating function
\begin{equation}
\label{eq:momgenmultsub}
\Lambda(\boldsymbol{\lambda})={\rm Tr}\left\{e^{-i\frac{\lambda_N}{2}\hat{A}_N(\tau_N)}\cdots e^{-i\frac{\lambda_1}{2}\hat{A}_1(\tau_1)}\hat{\rho}_0e^{-i\frac{\lambda_1}{2}\hat{A}_1(\tau_1)}\cdots e^{-i\frac{\lambda_N}{2}\hat{A}_N(\tau_N)} \right\}.
\end{equation}
The apparent symmetry between terms on the left and on the right of the density matrix is equivalent to treating the forward- and the backward-in-time branch of the time evolution symmetrically and determines the time-ordering of operators when evaluating higher moments (see below).

%

\subsection{Examples}
We now introduce two well-known examples of the KQPD: the Wigner function and the Full counting statistics (FCS). To recover the Wigner function from the moment generating function in Eq.~\eqref{eq:kmomgen}, we set
\begin{equation}
\label{eq:lamsetwig}
\sum_lf_l(\tau)\lambda_l\hat{A}_l(\tau)=\delta(\tau)\left[\lambda_x\hat{x}(\tau)+\lambda_p\hat{p}(\tau)\right],
\end{equation}
which results in
\begin{equation}
\label{eq:wigmom}
\Lambda(\lambda_x,\lambda_p)={\rm Tr}\left\{e^{-\frac{i}{2}(\lambda_x\hat{x}+\lambda_p\hat{p})}\hat{\rho}_0e^{-\frac{i}{2}(\lambda_x\hat{x}+\lambda_p\hat{p})}\right\},
\end{equation}
where we identified $\hat{x}=\hat{x}(0)$ and similarly for $\hat{p}$. It is straightforward to verify that the Wigner function is indeed given by
\begin{equation}
\label{eq:wigner}
\mathcal{W}(x,p)=\frac{1}{(2\pi)^2}\int d\lambda_xd\lambda_p e^{i\lambda_x x+i\lambda_p p} \Lambda(\lambda_x,\lambda_p)=\frac{1}{2\pi}\int d\lambda_pe^{-i\lambda_p p}\left\langle x+\frac{\lambda_p}{2}\right|\hat{\rho}_0\left|x-\frac{\lambda_p}{2}\right\rangle.
\end{equation}
The Wigner function is thus recovered as the KQPD corresponding to a simultaneous observation of $\hat{x}$ and $\hat{p}$. In the next section, we discuss how the Wigner function can be used to predict a simultaneous (but imprecise) measurement of position and momentum \cite{arthurs:1964}. The moments obtained from the Wigner function
\begin{equation}
\label{eq:momwig}
\langle x^n p^m\rangle_W=\int dx dp \,x^np^m \mathcal{W}(x,p),
\end{equation}
are determined by the Weyl order \cite{carmichael:book} obtained by complete symmetrization, i.e. every possible order is given the same weight, e.g.
\begin{equation}
\label{eq:weylord}
\langle x p\rangle_W=\frac{1}{2}{\rm Tr}\left\{(\hat{x}\hat{p}+\hat{p}\hat{x})\hat{\rho}_0\right\},\hspace{1cm}
\langle x^2p\rangle_W=\frac{1}{3}{\rm Tr}\left\{(\hat{x}^2\hat{p}+\hat{x}\hat{p}\hat{x}+\hat{p}\hat{x}^2)\hat{\rho}_0\right\}.
\end{equation}
This is a general result for the KQPD; when considering multiple operators at equal times, the moments generated by the KQPD are determined by the quantum averages of the completely symmetrized operators.

The next example we consider in this section is the FCS which is obtained from the moment generating function in Eq.~\eqref{eq:kmomgen} upon setting
\begin{equation}
\label{eq:lamsetfcs}
\sum_lf_l(\tau)\lambda_l\hat{A}_l(\tau)=\lambda \hat{A}(\tau),
\end{equation}
which results in
\begin{equation}
\label{eq:momgenfcs}
	\Lambda(\lambda)={\rm Tr}\left\{\hat{\mathcal{T}}_+e^{-i\frac{\lambda}{2}\int_{0}^{\mathcal{T}}d\tau\hat{A}(\tau)}\hat{\rho}_0 \hat{\mathcal{T}}_-e^{-i\frac{\lambda}{2}\int_{0}^{\mathcal{T}}d\tau\hat{A}(\tau)}\right\}={\rm Tr}\left\{\hat{\mathcal{T}}_Ke^{-i\frac{\lambda}{2}\int_Kd\tau\hat{A}(\tau)}\hat{\rho}_0\right\}.
\end{equation}
The FCS is given by the Fourier transform of the last expression
\begin{equation}
\label{eq:fcs}
\mathcal{F}(m)\equiv\frac{1}{2\pi}\int d\lambda e^{i\lambda m}\Lambda(\lambda),
\end{equation}
and gives information on the fluctuations of the time integral of $\hat{A}$. The FCS has been investigated intensively in the context of electron transport \cite{nazarov:book2}, where $\hat{A}$ corresponds to the current operator and one is interested in the fluctuations of the charge that is transferred through a conductor during a time interval $[0,\mathcal{T}]$.

The moments of the FCS can conveniently be expressed using the Keldysh contour
\begin{equation}
\label{eq:momentsfcs}
\int dm\, m^k\mathcal{F}(m)=i^k\partial_\lambda^k\Lambda(\lambda)\big|_{\lambda=0}=\int_Kd\tau_1\cdots d\tau_k{\rm Tr}\left\{ \hat{\mathcal{T}}_K\hat{A}(\tau_1)\cdots \hat{A}(\tau_k) \hat{\rho}_0\right\}.
\end{equation}
We thus find that the time-ordering corresponding to the KQPD is determined by the \textit{Keldysh time-ordering}, implying that operators are ordered along the Keldysh contour.
The Wigner function as well as the FCS have been discussed extensively in the literature. Explicit evaluations exhibiting negativity can be found in, e.g., Refs.~\cite{kenfack:2004,hofer:2016}.

\begin{figure}[t]
\centering
\includegraphics[width=.6\textwidth]{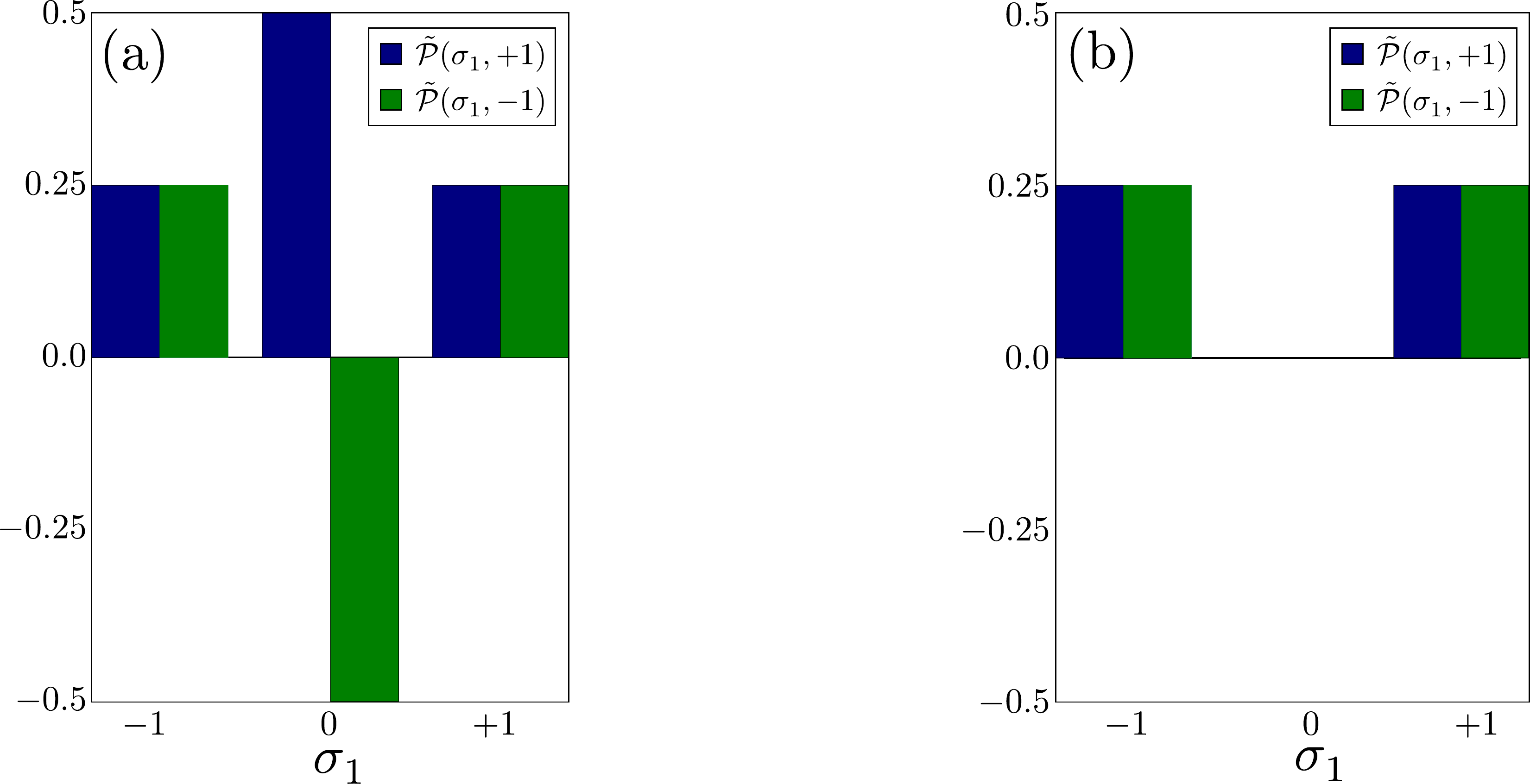}
\caption{Keldysh quasi-probability distribution for two subsequent spin measurements on a pure qubit state. While the second argument is restricted to the eigenvalues of $\hat{\sigma}_2$, the first argument can also be zero. (a) $\hat{\sigma}_0=\hat{\sigma}_2=\hat{\sigma}_x$ and $\hat{\sigma}_1=\hat{\sigma}_z$. In this case, a measurement of $\hat{\sigma}_1$ influences a subsequent measurement of $\hat{\sigma}_2$ which results in a negative value for $\tilde{\mathcal{P}}(0,-1)$. (b) $\hat{\sigma}_0=\hat{\sigma}_x$, $\hat{\sigma}_1=\hat{\sigma}_y$, and $\hat{\sigma}_2=\sigma_z$. In this case, the measurement outcome of $\hat{\sigma}_2$ is completely random, no matter if the first measurement is performed or not. A measurement of $\hat{\sigma}_1$ does therefore not influence a subsequent measurement of $\hat{\sigma}_2$ and the KQPD is purely positive.}
  \label{fig:barchart}
\end{figure}

To further illustrate the KQPD, we consider the very simple example of two subsequent spin measurements. The operators of interest are given by the rotated Pauli matrices $\hat{\sigma}_1$ and $\hat{\sigma}_2$. For simplicity, we consider a qubit in a pure state $|+_0\rangle$ which is an eigenstate of $\hat{\sigma}_0$ with eigenvalue $+1$. The KQPD then reads
\begin{equation}
\label{eq:kqpdspins}
\mathcal{P}(\Sigma_1,\Sigma_2)=\frac{1}{(2\pi)^2}\int d\lambda_1d\lambda_2e^{i\lambda_1\Sigma_1+i\lambda_2\Sigma_2}{\rm Tr}\left\{e^{-i\frac{\lambda_2}{2}\hat{\sigma}_2}e^{-i\frac{\lambda_1}{2}\hat{\sigma}_1}|+_0\rangle \langle +_0|e^{-i\frac{\lambda_1}{2}\hat{\sigma}_1}e^{-i\frac{\lambda_2}{2}\hat{\sigma}_2}\right\}.
\end{equation}
Writing the eigenstates of $\hat{\sigma}_j$ in the eigenbasis of $\hat{\sigma}_1$
\begin{equation}
\label{eq:basis}
|+_0\rangle = \alpha|+_1\rangle +\beta|-_1\rangle\hspace{2cm}|+_2\rangle = \gamma|+_1\rangle +\delta|-_1\rangle,
\end{equation}
we can evaluate Eq.~\eqref{eq:kqpdspins}, which can be written as
\begin{equation}
\mathcal{P}(\Sigma_1,\Sigma_2)=\sum_{\sigma_1=0,\pm 1}\sum_{\sigma_2=\pm 1}\tilde{\mathcal{P}}(\sigma_1,\sigma_2)\delta(\Sigma_1-\sigma_1)\delta(\Sigma_2-\sigma_2),
\end{equation}
with the discrete probability distribution
\begin{equation}
\label{eq:kqpdspinsdisc}
\begin{aligned}
&\tilde{\mathcal{P}}(+1,+1)=|\alpha|^2|\gamma|^2,\hspace{1cm}\tilde{\mathcal{P}}(-1,+1)=|\beta|^2|\delta|^2,\hspace{1cm}\tilde{\mathcal{P}}(0,+1)=2{\rm Re}\left\{\alpha\beta^*\gamma^*\delta\right\},\\&
\tilde{\mathcal{P}}(+1,-1)=|\alpha|^2|\delta|^2,\hspace{1cm}\tilde{\mathcal{P}}(-1,-1)=|\beta|^2|\gamma|^2,\hspace{1cm}\tilde{\mathcal{P}}(0,-1)=-2{\rm Re}\left\{\alpha\beta^*\gamma^*\delta\right\}.
\end{aligned}
\end{equation}
Note that although the operators $\hat{\sigma}_j$ have eigenvalues $\pm 1$, the KQPD is non-zero also for $\sigma_1=0$. Furthermore, these terms can become negative. We will discuss the origin of these terms and their negativity in more detail in the next section. For now, we only mention that the negativity reflects the fact that a measurement of $\hat{\sigma}_1$ disturbs a subsequent measurement of $\hat{\sigma}_2$. The negative terms thus vanish if these two operators are simultaneously measurable, $[\hat{\sigma}_1,\hat{\sigma}_2]=0$, if the first measurement does not influence the initial state, $[\hat{\sigma}_0,\hat{\sigma}_1]=0$, or if a measurement of $\hat{\sigma}_2$ has a completely random outcome no matter if $\hat{\sigma}_1$ is measured or not. The latter is the case if $\hat{\sigma}_0=\hat{\sigma}_x$, $\hat{\sigma}_1=\hat{\sigma}_y$, and $\hat{\sigma}_2=\sigma_z$. In all these cases, a measurement of $\hat{\sigma}_2$ yields outcomes which are independent of the presence or absence of a previous measurement of $\hat{\sigma}_1$. The KQPD for two subsequent spin measurements is illustrated in Fig.~\ref{fig:barchart}.

\subsection{Negativity}
\label{sec:negativity}
{As mentioned above, negativity in the KQPD indicates that measuring the considered observables necessarily influences the outcomes (i.e. measuring all of the operators results in different statistics than measuring each of them in a different realization). Following Ref.~\cite{hofer:2016}, we corroborate this claim by connecting negative values in the KQPD to coherent superpositions of different eigenstates of the observables. In the presence of such superpositions, any measurement (partly) collapses the state, possibly influencing subsequent measurement outcomes.} For clarity, we consider the case of $N$ observables at subsequent times as described by the moment generating function in Eq.~\eqref{eq:momgenmultsub}.
Assuming that the eigenstates of each operator $\hat{A}_l$ span the whole Hilbert space, we replace
\begin{equation}
\label{eq:exprepl}
e^{-i\frac{\lambda_l}{2}\hat{A}_l(\tau_l)}=\sum_{A_l^\alpha}e^{-i\frac{\lambda_l}{2}A_{l}^\alpha}\hat{U}(0,\tau_l)|A_{l}^{\alpha}\rangle\langle A_{l}^{\alpha}|\hat{U}(\tau_l,0),
\end{equation} 
where $\hat{A}_l|A_{l}^\alpha\rangle=A_{l}^\alpha|A_{l}^\alpha\rangle$ and the superscript $\alpha=L,R$ labels if the replacement happened on the left ($L$) or on the right ($R$) of the density matrix in Eq.~\eqref{eq:momgenmultsub}, i.e. $\alpha$ labels the branch (forward- or backward-in-time).
This allows us to take the $\lambda$-dependent factors out of the trace and obtain the KQPD by a Fourier transformation
\begin{equation}
\label{eq:kqpdtraj}
\mathcal{P}(\boldsymbol{A})=\sum_{\boldsymbol{A}_L,\boldsymbol{A}_R}\delta_{A_N^L,A_N^R}\delta\left(\boldsymbol{A}-\frac{\boldsymbol{A}_L+\boldsymbol{A}_R}{2}\right)\mathcal{A}(\boldsymbol{A}_L)\mathcal{A}^*(\boldsymbol{A}_R)\langle A^L_1|\hat{\rho}_{\tau_1}|A^R_1\rangle.
\end{equation}
Here the vectors $\boldsymbol{A}_{L/R}$ have as their entries $A_l^{L/R}$ and thus denote a particular manifestation of eigenvalues that we call a trajectory. The amplitudes associated with such a trajectory read
\begin{equation}
\label{eq:aampl}
\mathcal{A}(\boldsymbol{A}_\alpha)=\langle A_N^\alpha|\hat{U}(\tau_N,\tau_{N-1})|A_{N-1}^\alpha\rangle\langle A_{N-1}^\alpha|\cdots|A_{2}^\alpha\rangle\langle A_{2}^\alpha| \hat{U}(\tau_2,\tau_1)|A_1^\alpha\rangle,
\end{equation}
the Dirac delta for vectors is defined as
\begin{equation}
\delta\left(\boldsymbol{A}-\frac{\boldsymbol{A}_L+\boldsymbol{A}_R}{2}\right)=\prod_l \delta\left(A_l-\frac{A_l^R+A_l^L}{2}\right),
\end{equation}
and $\hat{\rho}_{\tau_1}=\hat{U}(\tau_1,0)\hat{\rho}_0\hat{U}(0,\tau_1)$.

Equation \eqref{eq:kqpdtraj} expresses the KQPD as a sum over pairs of trajectories. These pairs, here denoted left ($L$) and right ($R$) trajectory, are equivalent to a single trajectory along the Keldysh contour. A given pair of trajectories contributes with a weight that is given by the interference term $\mathcal{A}(\boldsymbol{A}_L)\mathcal{A}^*(\boldsymbol{A}_R)$ times a density matrix element $\langle A^L_1|\hat{\rho}_{\tau_1}|A^R_1\rangle$ corresponding to the ``starting point'' of the trajectories. Each pair contributes to the KQPD at the argument which corresponds to the \textit{mean} value of the observable on the two trajectories, again underlining the fact that the KQPD considers the fluctuations of the \textit{classical} field in the Keldysh language (see below). Finally, the Kronecker delta in Eq.~\eqref{eq:kqpdtraj} results from the trace and reflects the fact that the disturbance of the last measurement is irrelevant.

From Eq.~\eqref{eq:kqpdtraj}, we see that as long as $\boldsymbol{A}_L=\boldsymbol{A}_R$, the KQPD is strictly greater or equal to zero. It is therefore the interference terms of trajectories that differ from each other which allow the KQPD to become negative. This is a genuine quantum effect since these interference terms are a direct result of a coherent superposition between different eigenstates of the observables $\hat{A}_l$. {Although superposition can be encountered in classical (wave) theories, a coherent superposition between two states which correspond to two different values for an observable does not exist in classical theories (indeed, the presence of such superpositions violates the assumptions made by Leggett and Garg to obtain their inequalities \cite{leggett:1985}, see Sec.~\ref{eq:seclg})}.

In analogy to Eqs.~\eqref{eq:keldyshrot} and \eqref{eq:probapathint}, we can introduce a Keldysh rotation
\begin{equation}
\label{eq:keldyshrot2}
\boldsymbol{A}_{cl}=\frac{1}{2}(\boldsymbol{A}_L+\boldsymbol{A}_R),\hspace{2cm} \boldsymbol{A}_q = \boldsymbol{A}_L-\boldsymbol{A}_R,
\end{equation}
and write the KQPD in Eq.~\eqref{eq:kqpdtraj} as a path integral
\begin{equation}
\label{eq:probapathint2}
\mathcal{P}(\boldsymbol{A})=\int \mathcal{D}[\boldsymbol{A}_{cl},\boldsymbol{A}_q]\delta(\boldsymbol{A}-\boldsymbol{A}_{cl})e^{iS[\boldsymbol{A}_{cl},\boldsymbol{A}_q]},
\end{equation}
with the action
\begin{equation}
\label{eq:action}
e^{iS[\boldsymbol{A}_{cl},\boldsymbol{A}_q]}=\mathcal{A}(\boldsymbol{A}_{cl}+\boldsymbol{A}_q/2)\mathcal{A}^*(\boldsymbol{A}_{cl}-\boldsymbol{A}_q/2)\langle A^{cl}_1+A^q_1/2|\hat{\rho}_{\tau_1}| A^{cl}_1-A^q_1/2\rangle,
\end{equation}
and the integration measure
$\int \mathcal{D}[\boldsymbol{A}_{cl},\boldsymbol{A}_q]=\sum_{\boldsymbol{A}_{cl},\boldsymbol{A}_q}\delta_{A_N^q,0}$. In this language, the KQPD is given by the sum over all paths for which $\boldsymbol{A}=\boldsymbol{A}_{cl}$, weighted by the complex exponential of the action. From Eqs.~\eqref{eq:probapathint2} and \eqref{eq:action}, we see that any negativity in the KQPD necessarily implies the presence of a non-zero \textit{quantum} field $\boldsymbol{A}_q\neq0$.
In contrast to Eq.~\eqref{eq:probapathint}, which describes a single observable, the quantum field in Eq.~\eqref{eq:probapathint2} is not necessarily zero implying that the statistics of $\boldsymbol{A}_{cl}$, determined by the Keldysh time-ordering, are not necessarily equal to the statistics of $\boldsymbol{A}_L$ or $\boldsymbol{A}_R$. 

We now illustrate these ideas with the KQPD for two subsequent qubit measurements introduced in the last section. In terms of trajectories, the discrete KQPD in Eq.~\eqref{eq:kqpdspinsdisc} can be expressed as
\begin{equation}
\label{eq:kqpdtwospintraj}
\tilde{\mathcal{P}}(\sigma_1,\sigma_2)=\sum_{\sigma_1^{L/R}=\pm }\delta_{\sigma_1,(\sigma_1^L+\sigma_1^R)/2}\langle\sigma_2|\sigma_1^L\rangle\langle \sigma_1^L|+_0\rangle\langle +_0|\sigma_1^R\rangle\langle\sigma_1^R|\sigma_2\rangle.
\end{equation}
We thus see that the terms where $\sigma_1^L\neq\sigma_1^R$ are responsible for the terms $\tilde{\mathcal{P}}(0,\sigma_2)$. Note that these terms are only non-zero if the initial state is in a coherent superposition of the two eigenstates of $\hat{\sigma}_1$. One can further show that
\begin{equation}
\sum_{\sigma_1,\sigma_2=\pm 1}\tilde{\mathcal{P}}(\sigma_1,\sigma_2)=1,
\end{equation}
which directly implies that $\tilde{\mathcal{P}}(0,+1)=-\tilde{\mathcal{P}}(0,-1)$. Thus any contribution to Eq.~\eqref{eq:kqpdtwospintraj} from pairs of trajectories with $\sigma_1^L\neq\sigma_1^R$ results in negativity in the KQPD as illustrated in Fig.~\ref{fig:barchart}.

{We close this section with a brief comparison to other non-classicality indicators based on negative quasi-probabilities. As mentioned above, the Wigner function (at time $t$) is the KQPD for the observables $\hat{x}(t)$ and $\hat{p}(t)$. A negative Wigner function implies an impossibility of describing the corresponding state classically in the canonical position-momentum phase space (i.e. as a mixture of states with well defined positions and momenta). The KQPD generalizes this concept to arbitrary observables. Negativity in the KQPD implies that it is impossible to describe the system and its time evolution as a mixture of states with well defined values for the observable outcomes. We note that negative values in a discrete analogue of the Wigner function have recently been connected to the computational speed-up of a quantum computer \cite{veitch:2012,veitch:2013,howard:2014}.

In quantum optics, one often considers negativity, or strong singularity, in the Glauber-Sudarshan $P$ function as an indicator of non-classicality \cite{titulaer:1965,mandel:1986}. This criterion captures all states with non-positive Wigner functions as well as squeezed states and entangled states \cite{sperling:2009}. A non-classical $P$ function implies that the state can not be interpreted as a mixture of coherent states which indicates that the state can not be fully described by classical optics \cite{mandel:1986}. We note that the $P$ function has recently been extended to multiple times \cite{vogel:2008,krumm:2016,krumm:2017}. In contrast to the KQPD, this distribution is relevant for photo-detection and its moments correspond to normal ordering and regular time ordering.

While these QPDs only contain information on the states and possibly their time-evolution, quasi-probability \textit{representations} also describe measurements with QPDs \cite{spekkens:2008,ferrie:2008,ferrie:2009,ferrie:2011}. It has been shown that a positive representation (i.e. a representation that only uses positive probability distributions) of all states and measurements is impossible. Restricting the states and measurements to a given experimental procedure, a procedure is then said to require non-classical resources if no positive representation can be found \cite{ferrie:2008}. It is an interesting question if negativity in the KQPD implies the absence of a positive representation for a corresponding experimental procedure.}

\section{von Neumann measurements}
\label{sec:vonneumann}
In this section, we consider von Neumann type measurements and we show that the corresponding probability distributions can be obtained by convolving the KQPD with the back-action and imprecision associated with the measurement. We start by considering the instantaneous measurement of a single observable before moving on to subsequent instantaneous measurements, simultaneous measurements, and measurements of finite duration.

\subsection{Instantaneous measurements}
Following von Neumann \cite{vonneumann:book}, we model the measurement of an observable $\hat{A}$ at time $\tau$ by coupling it to a detector described by the canonically conjugate observables $\hat{r}$ and $\hat{\pi}$ via the interaction Hamiltonian
\begin{equation}
\label{eq:hvn}
\hat{H}_m = \delta(t-\tau)\chi\hat{A}\hat{\pi},
\end{equation}
where $\chi$ denotes the coupling strength. This interaction induces the time-evolution
$\hat{U}_m=\exp(-i\chi\hat{A}\hat{\pi})$,
which displaces the detector coordinate $\hat{r}$ by an amount that depends on the state of the system. Projectively measuring the position of the detector then yields the measurement of $\hat{A}$. Measuring a single observable in this way yields the probability distribution (see App.~\ref{app:vonneumann})
\begin{equation}
\label{eq:probaone}
P(A)=\int \chi dA'\langle  \chi(A-A')|\hat{\rho}_{\rm det}|\chi(A-A')\rangle\mathcal{P}(A';\tau),
\end{equation}
where $\hat{\rho}_{\rm det}$ is the density matrix of the detector, $|\chi(A-A')\rangle$ is an eigenstate of the operator $\hat{r}$, and $\mathcal{P}(A';\tau)$ is the KQPD for a single observable given in Eq.~\eqref{eq:probaa1}. Note that the position uncertainty in the initial detector state directly translates into a measurement uncertainty. A projective measurement can be obtained by choosing a position eigenstate as the initial state for the detector. In this case, the measured distribution coincides with the KQPD. This is due to the fact that the back-action of the measurement is irrelevant for a single observable since we are not interested in the post-measured state.

For multiple subsequent (instantaneous) measurements, we couple each observable to a different detector. The measurement Hamiltonian then reads 
\begin{equation}
\label{eq:hvnm}
\hat{H}_m = \sum_{l=1}^N\delta(t-\tau_l)\chi_l\hat{A}_l\hat{\pi}_l.
\end{equation}
In contrast to a single observable, each measurement can now influence all subsequent measurements. To capture this back-action effect, we evaluate the KQPD with the additional term $\sum_l\delta(t-\tau_l)\gamma_l\hat{A}_l$ in the Hamiltonian,
where the $\gamma_j$ are random variables which are distributed according to the detector states (see below). This yields the modified moment generating function
\begin{equation}
\label{eq:momgenmultsubba}
\Lambda(\boldsymbol{\lambda};\boldsymbol{\gamma})={\rm Tr}\left\{e^{-i\left(\frac{\lambda_N}{2}+\gamma_N\right)\hat{A}_N(\tau_N)}\cdots e^{-i\left(\frac{\lambda_1}{2}+\gamma_1\right)\hat{A}_1(\tau_1)}\hat{\rho}_0e^{-i\left(\frac{\lambda_1}{2}-\gamma_1\right)\hat{A}_1(\tau_1)}\cdots e^{-i\left(\frac{\lambda_N}{2}-\gamma_N\right)\hat{A}_N(\tau_N)} \right\},
\end{equation}
which can be obtained by replacing $\lambda_l\rightarrow \lambda_l\pm 2\gamma_l$ in Eq.~\eqref{eq:momgenmultsub}, where the $+$ ($-$) sign corresponds to terms on the left (right) side of the density matrix.

The probability distribution which describes subsequent von Neumann measurements can then be shown to read (for a derivation, see App.~\ref{app:vonneumann})
\begin{equation}
\label{eq:probasub}
P(\boldsymbol{A})=\int d\boldsymbol{A}'d\boldsymbol{\gamma}\left[\prod_l \mathcal{W}_l(\chi_l[A_l-A_l'],\gamma_l/\chi_l)\right]\mathcal{P}(\boldsymbol{A}';\boldsymbol{\gamma}),
\end{equation}
where $\mathcal{W}_l(r,\gamma)$ denotes the Wigner function of detector $l$ and $\mathcal{P}(\boldsymbol{A}';\boldsymbol{\gamma})$ is the KQPD defined as the Fourier transform of the moment generating function in Eq.~\eqref{eq:momgenmultsubba}. The measurement outcomes are determined by a convolution of the KQPD and the Wigner functions of the detectors. We can thus interpret the KQPD as describing the \textit{intrinsic} fluctuations of the observables of interest which is being influenced by the effects of the measurement \cite{clerk:2011}. The position distributions of the detectors determine the precision of the measurement while the momentum distributions determine the back-action via the $\gamma_l$-dependent terms in the KQPD. The Heisenberg uncertainty relation implies a trade-off between these two effects, indicating that we can never eliminate both influences.

As an example, we consider a measurement of position $\hat{x}$ at $\tau=0$ followed immediately by a momentum $\hat{p}$ measurement. The probability distribution describing such a measurement reads
\begin{equation}
\label{eq:probaxp}
W_{x\rightarrow p}(x,p)=\int dx'dp'd\gamma_xd\gamma_p\mathcal{W}_p(\chi_p[p-p'],\gamma_p/\chi_p)\mathcal{W}_x(\chi_x[x-x'],\gamma_x/\chi_x)\mathcal{W}(x',p'+\gamma_x).
\end{equation} 
Here $\mathcal{W}_{x/p}$ denotes the Wigner function of the detector measuring $x/p$ and $\mathcal{W}$ without a subscript denotes the KQPD which in this case is given by the Wigner function of the system. The position measurement disturbs the momentum of the system as reflected by the second argument of the KQPD. Note that it is the momentum distribution of the $x$-detector which determines the effect of this disturbance. This reflects the trade-off between the precision of the position measurement (which requires a peaked position distribution for the $x$-detector) and the disturbance of the subsequent momentum distribution (which is minimized for a peaked momentum distribution for the $x$-detector). Note that the last expression is independent of the momentum distribution of the $p$-detector (this can be seen by performing the integral over $\gamma_p$), reflecting the fact that the back-action of the momentum measurement is irrelevant since it is the last measurement. Inverting the order of the measurements yields a probability distribution which is given by the last expression upon exchanging $\mathcal{W}(x',p'+\gamma_x)\rightarrow \mathcal{W}(x'-\gamma_p,p')$.

Returning to arbitrary observables and choosing Gaussian Wigner functions for the detectors
\begin{equation}
\mathcal{W}_l(r,\gamma)=\frac{1}{2\pi\sigma_{r_l}\sigma_{\gamma_l}}e^{-\frac{r^2}{2\sigma_{r_l}^2}}e^{-\frac{\gamma^2}{2\sigma_{\gamma_l}^2}},
\end{equation}
we can write the probability distribution in Eq.~\eqref{eq:probasub} in terms of trajectories [cf.~Eq.~\eqref{eq:kqpdtraj}] yielding
\begin{equation}
\label{eq:probasubtraj}
\begin{aligned}
P(\boldsymbol{A})=\sum_{\boldsymbol{A}_L,\boldsymbol{A}_R}\delta_{A_N^L,A_N^R}&\left[\prod_l\frac{1}{\sqrt{2\pi}\sigma_{{\rm imp},l}}e^{-\frac{1}{2\sigma_{{\rm imp},l}^2}\left(A_l-\frac{A_l^L+A_l^R}{2}\right)^2}e^{-\frac{\sigma_{{\rm BA},l}^2}{2}\left(A_l^L-A_l^R\right)^2}\right]\\&\times\mathcal{A}(\boldsymbol{A}_L)\mathcal{A}^*(\boldsymbol{A}_R)\langle A^L_1|\hat{\rho}_{\tau_1}|A^R_1\rangle.
\end{aligned},
\end{equation}
where $\sigma_{{\rm imp},l}=\sigma_{r_l}/\chi_l$ and $\sigma_{{\rm BA},l}=\sigma_{\gamma_l}\chi_l$. Comparing the last expression with the KQPD in Eq.~\eqref{eq:kqpdtraj}, we can clearly identify the effects of the measurement imprecision and back-action. The first Gaussian in the last expression replaces the Dirac delta in Eq.~\eqref{eq:kqpdtraj}. It implies that even terms with $A_l\neq (A_l^L+A_l^R)/2$ contribute to the probability of measuring $A_l$. This effect captures the imprecision of the measurement which becomes more pronounced as the uncertainty in the position of detector $l$ grows. The second Gaussian in the last expression reduces the weight of terms which have $A_l^L\neq A_l^R$, i.e. a finite quantum field in the Keldysh language. {This effect is due to the measurement back-action which (partly) collapses the coherent superpositions between eigenstates of $\hat{A}_l$}. The coupling Hamiltonian in Eq.~\eqref{eq:hvnm} entangles the system with the detector leading to dephasing in the reduced state of the system. As the uncertainty in the detector momentum grows, this effect becomes more pronounced. From the Heisenberg uncertainty relation, we find
\begin{equation}
\label{eq:heisenberg}
\sigma_{{\rm imp},l}\sigma_{{\rm BA},l}=\sigma_{r_l}\sigma_{\gamma_l}\geq\frac{1}{2},
\end{equation}
reflecting the trade-off between measurement imprecision and back-action {(see also Ref.~\cite{clerk:2010})} which ensures that the measured distribution is positive even if the KQPD exhibits negativities. We note that coupling the system to additional degrees of freedom, such as a thermal reservoir, introduces dephasing analogously to the measurement back-action \cite{hofer:2016}.

\begin{figure}[t]
\centering
\includegraphics[width=.8\textwidth]{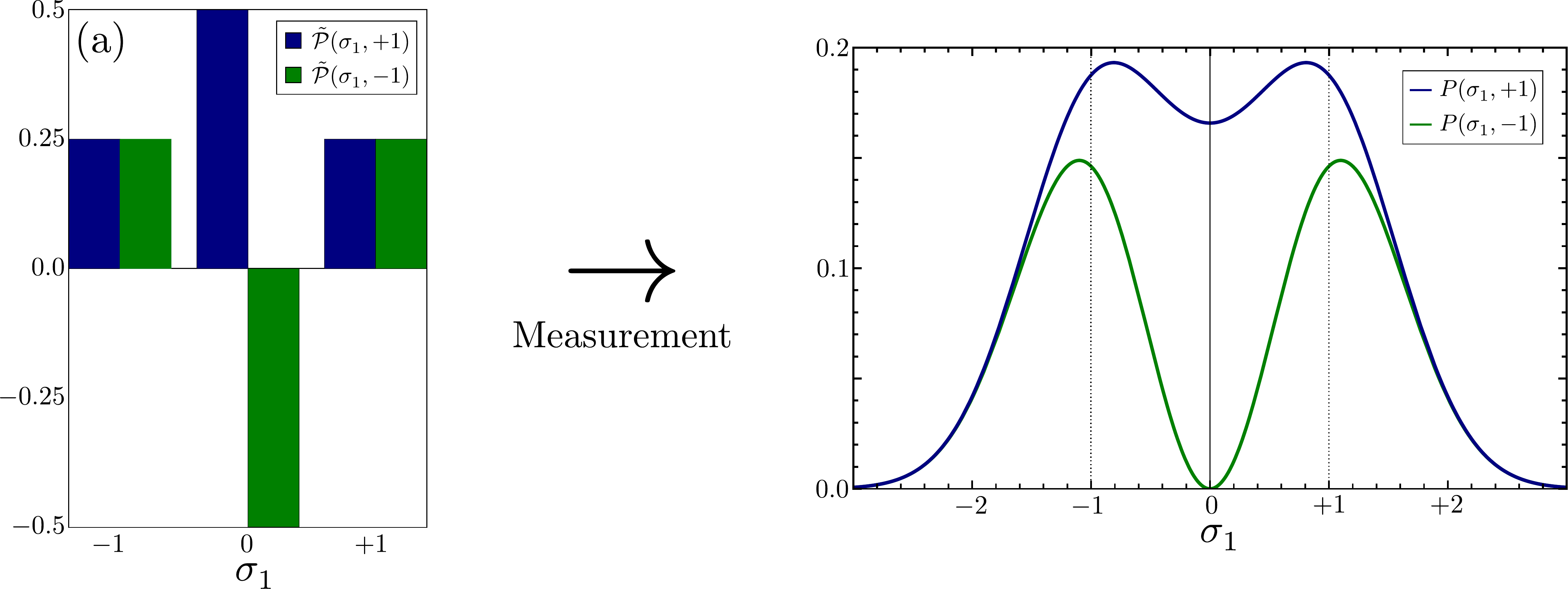}
\caption{Effect of measurement imprecision and back-action on the KQPD. Here $\hat{\sigma}_0=\hat{\sigma}_2=\hat{\sigma}_x$ and $\hat{\sigma}_1=\hat{\sigma}_z$ to maximize the negativity in the KQPD. The measurement imprecision and back-action modify the KQPD resulting in a positive probability distribution. Here $\sigma_{\rm imp}\sigma_{\rm BA}=1/2$, minimizing the influence of the measurement. Depending on the measurement outcome of the second (projective) measurement, the distribution for the first measurement results in two clearly distinguishable outcomes or not. This is directly related to the terms $\tilde{\mathcal{P}}(0,\sigma_2)$ which are responsible for the negative values in the KQPD. Note that the peaks are shifted with respect to the eigenvalues of $\hat{\sigma}_1$ (dashed lines). This effect allows for the occurrence of anomalous weak values as discussed in Sec.~\ref{sec:weakvalues}.}
  \label{fig:measured}
\end{figure}

To illustrate the effect of the measurement, we return to our example of two subsequent spin measurements [cf.~Eq.~\eqref{eq:kqpdspins}]. Since the back-action of the second measurement is irrelevant, we consider a measurement of intermediate strength of $\hat{\sigma}_1$ followed by a projective measurement of $\hat{\sigma}_2$.
In terms of the discrete KQPD given in Eq.~\eqref{eq:kqpdspinsdisc}, we find for the distribution describing a von Neumann measurement (with Gaussian detectors)
\begin{equation}
\label{eq:kqpdspinsmeas}
P(\sigma_1,\sigma_2)=\frac{1}{\sqrt{2\pi}\sigma_{\rm imp}}\sum_{\sigma'_1=0,\pm 1}e^{-\frac{1}{2\sigma_{\rm imp}^2}(\sigma_1-\sigma_1')^2}e^{-2\sigma_{\rm BA}^2\delta_{\sigma'_1,0}}\tilde{\mathcal{P}}(\sigma_1',\sigma_2).
\end{equation}
The effect of the measurement is illustrated in Fig.~\ref{fig:measured}. We note that the distribution of measurement outcomes for the first measurement depends very strongly on the outcome of the second measurement. Specifically, we see from Fig.~\ref{fig:measured} that only considering events where the second (projective) measurement yields $-1$ allows one to distinguish the outcomes of the first measurement much better (for a given measurement strength). This effect, which arises from the negativity of the KQPD, is responsible for the occurrence of anomalous weak values as discussed in Sec.~\ref{sec:weakvalues}. 

{Note that subsequent von Neumann measurements can also be used to model continuous measurements, a subject which has received a lot of attention recently, see e.g. \cite{korotkov:1999,korotkov:2001,chantasri:2013,weber:2014,chantasri:2015}.}

\subsection{Simultaneous measurements of position and momentum}

We now turn to the simultaneous measurement of position and momentum. In analogy to Eq.~\eqref{eq:hvn} (and following Arthurs and Kelly \cite{arthurs:1964}), we model the measurement by the simultaneous coupling of two detectors to the position and momentum of the system respectively
\begin{equation}
\label{eq:hvnmsim}
\hat{H}_m = \delta(t)\left(\chi_x\hat{x}\hat{\pi}_x+\chi_p\hat{p}\hat{\pi}_p\right).
\end{equation}
It can be shown that this measurement is described by the probability distribution \cite{di_lorenzo:2011}
\begin{equation}
\label{eq:probaxpsim}
W(x,p)=\int dx'dp'd\gamma_xd\gamma_p\mathcal{W}_p(\chi_p[p-p'],\gamma_p/\chi_p)\mathcal{W}_x(\chi_x[x-x'],\gamma_x/\chi_x)\mathcal{W}(x'-\gamma_p/2,p'+\gamma_x/2).
\end{equation}
The arguments of the last Wigner function reflect the fact that in a simultaneous measurement, both position and momentum are being disturbed. Note the close similarity between the last expression and the probability distribution describing subsequent measurements of position and momentum [cf.~Eq.~\eqref{eq:probaxp}]. Notably, the KQPD is in both cases given by the Wigner function. This is a consequence of the commutation relation $[\hat{x},\hat{p}]=i$ and does not hold in general for non-commuting observables.

For a detailed discussion on the von Neumann type simultaneous measurement of position and momentum, we refer the reader to Ref.~\cite{stenholm:1992}. Here we only discuss the case of Gaussian states of minimal uncertainty for the detectors ($k=x,p$)
\begin{equation}
\mathcal{W}_{k}(r,\gamma)=\frac{1}{\pi}e^{-r^2/\sigma_{k}^2}e^{-\gamma^2\sigma_{k}^2},
\end{equation}
with the additional constraint $\sigma_x\sigma_p/(\chi_x\chi_p)=1/2$. This constrained minimizes the disturbance of the measurement and yields the Husimi $Q$-function as the measured probability distribution \cite{husimi:1940,stenholm:1992}
\begin{equation}
\label{eq:husimiq}
W(x,p)=\frac{1}{\pi}\int dx' dp' e^{-\frac{(x-x')^2}{2\sigma^2}}e^{-2\sigma^2(p-p')^2}\mathcal{W}(x',p'),
\end{equation}
where $\sigma=\sigma_x/\chi_x$. In addition to this von Neumann type measurement, the Husimi $Q$-function also describes heterodyne measurements of two field quadratures \cite{leonhardt:1995}.

\subsection{Measuring the FCS}


In this section, we consider a von Neumann type measurement of the time-integral of an observable \cite{nazarov:2003}. To this end, we couple a detector to the observable of interest for a finite amount of time
\begin{equation}
\label{eq:hvnc}
\hat{H}_m = \Theta(t)\Theta(\mathcal{T}-t)\chi\hat{A}\hat{\pi}.
\end{equation}
This measurement can be shown to be described by the probability distribution \cite{nazarov:2003}
\begin{equation}
\label{eq:measfcs}
F(m) = \int d\gamma \mathcal{W}(\chi[m-m'],\gamma/\chi)\mathcal{F}(m';\gamma),
\end{equation}
where $\mathcal{F}(m;\gamma)$ is obtained through Eqs.~\eqref{eq:momgenfcs} and \eqref{eq:fcs} upon adding the term $\Theta(t)\Theta(\mathcal{T}-t)\gamma\hat{A}$ to the Hamiltonian.
Again, we find a trade-off between measurement imprecision, depending on the spread in the detector position, and back-action which depends on the spread in the detector momentum. As discussed in Ref.~\cite{hofer:2016}, the last equation can be expressed in terms of trajectories yielding (for a Gaussian detector)
\begin{equation}
\label{eq:fcsmeastraj}
F(m)=\frac{1}{\sqrt{2\pi}\sigma_{\rm imp}}\sum_{\boldsymbol{A}_L,\boldsymbol{A}_R}\delta_{A_\mathcal{T}^L,A_\mathcal{T}^R}e^{-\frac{\left(m-\frac{1}{2}m_L-\frac{1}{2}m_R\right)^2}{2\sigma^2_{\rm imp}}} e^{-\frac{\sigma_{\rm BA}^2}{2}\left(m_L-m_R\right)^2}\langle A_0^L|\hat{\rho}|A_0^R\rangle \mathcal{A}(\boldsymbol{A}_L)\mathcal{A}^*(\boldsymbol{A}_R),
\end{equation}
where the trajectory amplitudes are obtained by dividing the time evolution into $N$ time-slices of infinitesimal duration $\delta t$ and inserting identities resolved in the eigenstates of $\hat{A}$ after each slice
\begin{equation}
\label{eq:aamplfcs}
\mathcal{A}(\boldsymbol{A}_\alpha)=\langle A_\mathcal{T}^\alpha|\hat{U}(\mathcal{T},\mathcal{T}-\delta t)|A_{\mathcal{T}-\delta t}^\alpha\rangle\langle A_{\tau-\delta t}^\alpha|\cdots|A_{\delta t}^\alpha\rangle\langle A_{\delta t}^\alpha| \hat{U}(\delta t,0)|A_0^\alpha\rangle.
\end{equation}
The discrete time-integral over such a trajectory is denoted by $m_\alpha =\sum_l A^\alpha_l \delta t$. The Heisenberg uncertainty relation applied to the detector variables again enforces $\sigma_{\rm imp}\sigma_{\rm BA}\geq 1/2$, reflecting the trade-off between smearing out the negativities in the FCS by measurement imprecision and reducing the weight of negative terms (which necessarily have $m_L\neq m_R$) by measurement back-action in complete analogy to Eq.~\eqref{eq:probasubtraj}.

This concludes our treatment on von Neumann measurements. Expressing the measured probability distributions in terms of the KQPD suggests that the latter should be interpreted as describing the intrinsic fluctuations of the observables of interest. In the next section, we will build on this interpretation and demonstrate the utility of studying the KQPD to understand the quantum behavior of dynamic systems.

\section{Applications of the KQPD}
\label{sec:applications}
We now turn to situations where it is beneficial to study the KQPD in the absence of any measurement device. We are especially interested in situations where the KQPD fails to be positive. As discussed in Sec.~\ref{sec:negativity}, this indicates non-classical behavior. The most prominent KQPD is by far the Wigner function, which is a representation of the density matrix and is used extensively to illustrate the state of a quantum system and to capture its non-classicality \cite{deleglise:2008,kenfack:2004}. An example for a KQPD which goes beyond instantaneous states is provided by the FCS which describes the fluctuations of an observable over a certain time-interval. While the FCS has mostly been employed to describe electronic transport in systems where it is a purely positive function, its utility for detecting non-classical behavior has been recognized \cite{hofer:2016,clerk:2011,bednorz:2010,bednorz:2012,belzig:2001,clerk:2010}. Here we discuss three additional applications of the KQPD, stressing its versatility and its utility in capturing non-classical behavior in the dynamics of quantum systems.

\subsection{Work fluctuations}

Since work is not a state function, it is in general not sufficient to interrogate a system at a single time in order to determine the work performed by the system. As we have seen above, performing measurements at multiple times generally results in a probability distribution which is influenced by the measurement itself. As a consequence, the observed amount of work will in general depend on the way it is measured \cite{perarnau:2017}. Here we use the KQPD to describe work as a measurement-independent, fluctuating quantity. We note that the KQPD has been used before to describe work fluctuations \cite{esposito:2009,solinas:2015,solinas:2016}. 

In classical statistical mechanics, work is defined as the energy change in the macroscopic, accessible degrees of freedom, while heat is the energy change in the microscopic, inaccessible degrees of freedom. In quantum systems, all degrees of freedom are in principle microscopic. We can however still define work as being the energy change mediated by the accessible degrees of freedom. Here we consider two different scenarios: In the first, one has access to the energy stored in a part of the system that we call the work storage device. The energy increase of this device is then interpreted as the work performed by the system. {This is the analog of determining the work performed on a suspended weight by measuring its height.} In the second scenario, one has access to the power produced by the system and the work is defined as the time-integral of the power. This captures the situation in thermoelectric heat engines, where one can access the power through a measurement of the electrical current. {In terms of the suspended weight, this corresponds to the analog of measuring the power by measuring the velocity of the weight.}

For the first scenario, we write the total Hamiltonian as $\hat{H}=\hat{H}_w+\hat{H}_{\rm rest}$, where $\hat{H}_w$ denotes the Hamiltonian of the work storage device. Work fluctuations can then be described by the KQPD corresponding to the difference of two subsequent energy measurements on the work storage device
\begin{equation}
\label{eq:work}
\begin{aligned}
\mathcal{P}(w;\tau)=&\frac{1}{2\pi}\int d\lambda e^{i\lambda w}{\rm Tr}\left\{e^{-i\frac{\lambda}{2}\hat{H}_w}\hat{U}(\tau,0)e^{i\frac{\lambda}{2}\hat{H}_w}\hat{\rho}_0e^{i\frac{\lambda}{2}\hat{H}_w}\hat{U}(0,\tau)e^{-i\frac{\lambda}{2}\hat{H}_w}\right\}\\
=&\sum_{E_\tau,E_L,E_R}\delta\left(w-\left[E_\tau-\frac{E_L+E_R}{2}\right]\right){\rm Tr}\left\{\langle E_\tau|\hat{U}(\tau,0)|E_L\rangle\langle E_L|\hat{\rho}_0|E_R\rangle \langle E_R|\hat{U}(0,\tau)|E_\tau\rangle\right\}.
\end{aligned}
\end{equation}
Here $\hat{H}_w|E_j\rangle=E_j|E_j\rangle$ and $\tau$ denotes the time interval between the two measurements. For initial states which commute with $\hat{H}_w$, we find $E_L=E_R$ and we recover the probability for obtaining $w$ as the difference of two projective energy measurements \cite{talkner:2007}. Any negativity in the above distribution is thus due to the coherence in the initial state. We note that the probability distribution obtained by performing two Gaussian energy measurements, as discussed in Ref.~\cite{talkner:2016}, is obtained from Eq.~\eqref{eq:work} by including the imprecision and back-action of the measurement as discussed in Sec.~\ref{sec:vonneumann}.
The moments generated by the KQPD are given by
\begin{equation}
\label{eq:momentswork}
\langle w^k\rangle = \sum_{l=0}^{k}\frac{1}{2^k}{k \choose l}{\rm Tr}\left\{ \hat{\mathcal{T}}_-\left(\hat{H}_w(\tau)-\hat{H}_w\right)^l\hat{\mathcal{T}}_+\left(\hat{H}_w(\tau)-\hat{H}_w\right)^{(k-l)}\hat{\rho}_0\right\},
\end{equation}
where $\hat{H}_w(\tau)=\hat{U}(0,\tau)\hat{H}_w\hat{U}(\tau,0)$. For the first two moments, i.e. $k=1,2$ in the last expression, this reduces to 
\begin{equation}
\label{eq:momentswork12}
\langle w^k\rangle={\rm Tr}\left\{ \left(\hat{H}_w(\tau)-\hat{H}_w\right)^k\hat{\rho}_0\right\},\hspace{1cm}{\rm for}\hspace{.2cm} k=1,2.
\end{equation}
The mean work obtained through the KQPD is thus just given by the difference of the average energy in the work storage device at $t=\tau$ and $t=0$. This is in contrast to the probability distribution obtained by performing two projective energy measurements, where the mean value only coincides with Eq.~\eqref{eq:momentswork12} for initial states that commute with $\hat{H}_w$. For an explicit evaluation of Eq.~\eqref{eq:work}, we refer to Ref.~\cite{solinas:2016}. See Ref.~\cite{batalhao:2014} for an experimental reconstruction of the work distribution by measuring its moment generating function (for an initial state that commutes with the Hamiltonian).

In case one has access to a power operator $\hat{P}$, the KQPD describing work fluctuations is given by the FCS of the power operator. The KQPD and its moments are then given by Eqs.~(\ref{eq:momgenfcs}-\ref{eq:momentsfcs}) upon replacing $\hat{A}$ with $\hat{P}$. {This scenario is relevant for thermoelectric devices, where the power operator is given by the electrical current operator times the voltage across the device $\hat{P}=\hat{I}V$. In this case, the work fluctuations reduce to the problem of charge FCS \cite{nazarov:book2}. We now return to systems described by $\hat{H}=\hat{H}_w+\hat{H}_{\rm rest}$, where we take the Hamiltonians to be time-independent for simplicity (see Ref.~\cite{venkatesh:2015} for a discussion on the power operator for a time-dependent Hamiltonian). In this case, the power operator is given by the energy change of the work storage device per unit time, i.e. $\hat{P}=i[\hat{H},\hat{H}_w]$.}
If the power operator commutes with the total Hamiltonian, i.e. $[\hat{P},\hat{H}]=0$, we find
\begin{equation}
\label{eq:powerop}
e^{-i\frac{\lambda}{2}\hat{H}_w}\hat{U}(\tau,0)e^{i\frac{\lambda}{2}\hat{H}_w}=\hat{\mathcal{T}}_+e^{-i\frac{\lambda}{2}\int_0^\tau dt \hat{P}(t)}.
\end{equation}
This implies that the FCS of the power operator is equal to the KQPD for two energy measurements given in Eq.~\eqref{eq:work} \cite{esposito:2009}. The difference of the two scenarios considered here is then only relevant when taking into account the disturbance from an actual measurement. {For a thermoelectric device, the energy in the work storage device corresponds to the electric potential energy stored in an electronic contact, i.e. $\hat{H}_w=V\hat{Q}$, where $\hat{Q}$ denotes the charge operator in the contact with $\hat{I}=i[\hat{H},\hat{Q}]$. The work provided by a thermoelectric device can thus either be determined by measuring the electrical current (determining the power) or by measuring the number of electrons in an electric contact (determining the energy in the work storage device). We note that both approaches have been considered in the theory of charge FCS \cite{nazarov:book2}.}

\subsection{Weak values}
\label{sec:weakvalues}
The weak value of an observable $\hat{A}$ is obtained by preparing the system in some state $|I\rangle$, performing a weak measurement of $\hat{A}$, and post-selecting the measurement outcomes on the final state $|F\rangle$ \cite{aharonov:1988}. Describing the weak measurement as a von Neumann measurement, the post-measured state of the \textit{detector} is determined by the weak value
\begin{equation}
\label{eq:weakvalue}
A_w=\frac{\langle F|\hat{A}|I\rangle}{\langle F|I\rangle},
\end{equation}
and the average of the measurement outcome is given by the real part of the last expression \cite{aharonov:1988,dressel:2015}. Any possible time-evolution between pre-selection, measurement and post-selection is absorbed in the definition of the initial and final states. In the KQPD framework, we can describe the weak value measurement using the distribution describing two subsequent measurements, one for the weak measurement and one for the post-selection. For simplicity, we do not treat the pre-selection as an additional measurement but impose it by setting $\hat{\rho}_0=|I\rangle\langle I|$. The relevant KQPD then reads
\begin{equation}
\label{eq:kqpdweak1}
\begin{aligned}
\mathcal{P}(A,B)=&\frac{1}{(2\pi)^2}\int d\lambda_Ad\lambda_Be^{i\lambda_A A}e^{i\lambda_B B}{\rm Tr}\left\{e^{-i\frac{\lambda_B}{2}\hat{B}}e^{-i\frac{\lambda_A}{2}\hat{A}}|I\rangle\langle I|e^{-i\frac{\lambda_A}{2}\hat{A}}e^{-i\frac{\lambda_B}{2}\hat{B}}\right\}\\
=&\sum_{B_F,A_L,A_R}\delta\left(B-B_F\right)\delta\left(A-\frac{A_L+A_R}{2}\right) \langle B_F|A_L\rangle\langle A_L|I\rangle\langle I|A_R\rangle\langle A_R|B_F\rangle,
\end{aligned}
\end{equation}
where $|B_F\rangle$ are eigenstates of $\hat{B}$. The KQPD corresponding to the measurement of the weak value is then obtained by taking into account the post-selection, i.e. fixing the argument corresponding to the second measurement
\begin{equation}
\label{eq:kqpdweak2}
\mathcal{P}_w(A)=\frac{\mathcal{P}(A,F)}{\int dA\mathcal{P}(A,F)}=\frac{1}{|\langle F|I\rangle|^2}\sum_{A_L,A_R}\delta\left(A-\frac{A_L+A_R}{2}\right)\langle F|A_L\rangle\langle A_L|I\rangle\langle I|A_R\rangle\langle A_R|F\rangle.
\end{equation}
The statistics of weak value measurements (with post-selection performed by a projective measurement) can be described using the last expression by including measurement imprecision and back-action as discussed in Sec.~\ref{sec:vonneumann}. In the weak measurement limit, the disturbance of the measurement does not alter the average value of the KQPD, which is equal to the real part of Eq.~\eqref{eq:weakvalue}
\begin{equation}
\label{eq:evgweakvalue}
\langle A\rangle_w =\int dA\,A\mathcal{P}_w(A)={\rm Re}\left\{\frac{\langle F|\hat{A}|I\rangle}{\langle F|I\rangle}\right\}.
\end{equation}

From Eq.~\eqref{eq:kqpdweak2}, we see that the KQPD vanishes for arguments that lie outside the spectrum of $\hat{A}$. However Eq.~\eqref{eq:evgweakvalue} can take on values which lie outside the spectrum of $\hat{A}$. In such cases, the weak value is called \textit{anomalous}.
We now show that an anomalous weak value implies that the distribution $\mathcal{P}_w(A)$ takes on negative values. To see this, assume that $\mathcal{P}_w(A)\geq 0$. In that case, we find
\begin{equation}
\label{eq:wvnegkqpd}
\langle A\rangle_w =\int_{A_{\rm min}}^{A_{\rm max}} dA\,A\mathcal{P}_w(A)\leq A_{\rm max} \int_{A_{\rm min}}^{A_{\rm max}} dA\mathcal{P}_w(A)=A_{\rm max},
\end{equation}
where $A_{\rm min \,(max)}$ is the smallest (largest) eigenvalue of $\hat{A}$ and a similar inequality restricts $\langle A\rangle_w\geq A_{\rm min}$. Note that the inequality in Eq.~\eqref{eq:wvnegkqpd} is only valid for a strictly positive KQPD.
An anomalous weak value thus necessarily implies that the KQPD $\mathcal{P}_w(A)$ [and also $\mathcal{P}(A,B)$, cf.~Eq.~\eqref{eq:kqpdweak2}] exhibits negativity indicating non-classical behavior. This is an agreement with Ref.~\cite{pusey:2014}, where anomalous weak values have been shown to imply contextuality. We note that weak values have been connected to quasi-probability distributions before \cite{dressel:2015}. In particular, the average momentum at a given position that is obtained from the Wigner function has been shown to be determined by the weak value of the momentum
\begin{equation}
\label{eq:weakvaluep}
\frac{\int dp\, p\mathcal{W}(x,p)}{\int dp \mathcal{W}(x,p)}={\rm Re}\left\{\frac{\langle x|\hat{p}|I\rangle}{\langle x|I\rangle}\right\},
\end{equation}
which follows from our approach when taking $\hat{A}=\hat{p}$ and $\hat{B}=\hat{x}$.

{Finally, we note that the distribution in Eq.~\eqref{eq:kqpdweak2} is relevant for a measurement of $\hat{A}$ with an arbitrary strength. It has the variance}
\begin{equation}
\label{eq:weakvariance}
\langle \hat{A}^2\rangle_w-\left(\langle \hat{A}\rangle_w\right)^2 = \frac{1}{2}{\rm Re}\left\{\frac{\langle F|\hat{A}^2|I\rangle}{\langle F|I\rangle}\right\}-\frac{1}{2}\left({\rm Re}\left\{\frac{\langle F|\hat{A}|I\rangle}{\langle F|I\rangle}\right\}\right)^2.
\end{equation}
For $\hat{A}=\hat{p}$ and $|F\rangle =|x\rangle$, the last expression is proportional to the \textit{quantum potential energy} \cite{dressel:2015}, a central object in Bohmian mechanics \cite{bohm:1952}. {See also Ref.~\cite{dressel:2012} for a connection between weak values and von Neumann measurements of arbitrary strength and post-selection.}

For our example of two subsequent spin measurements, we find [using Eqs.~\eqref{eq:kqpdspins}, \eqref{eq:kqpdspinsdisc}, and \eqref{eq:kqpdweak2}]
\begin{equation}
\label{eq:kqpdwvspins}
\mathcal{P}_w(\sigma_1)=\frac{|\alpha|^2|\delta|^2\delta(\sigma_1-1)+|\beta|^2|\gamma|^2\delta(\sigma_1+1)-2{\rm Re}\{\alpha\beta^*\gamma^*\delta\}\delta(\sigma_1)}{|\alpha\delta-\beta\gamma|^2},
\end{equation}
where we chose $|-_2\rangle$ as the final state $|F\rangle$.
We see that for a sufficiently small denominator, the probability density associated with $\sigma_1=\pm 1$ can become very large as long as the presence of the negative term ensures the normalization of $\mathcal{P}_w(\sigma_1)$. In this way, the weak value, determined by $\mathcal{P}_w(1)-\mathcal{P}_w(-1)$ can be amplified resulting in values far outside the range $[-1,+1]$. This can however only occur for a sufficiently large negative term in the KQPD as can be seen from Eq.~\eqref{eq:kqpdwvspins} which implies
\begin{equation}
\label{eq:weakvaluespins}
\langle \sigma_1\rangle_w>1\hspace{.5cm}\Leftrightarrow\hspace{.5cm} {\rm Re}\{\alpha\beta^*\gamma^*\delta\}>|\beta|^2|\gamma|^2.
\end{equation}

\subsection{Leggett-Garg inequalities}
\label{eq:seclg}

In Ref.~\cite{leggett:1985}, Leggett and Garg derived inequalities which hold under the assumption of \textit{macroscopic realism} [i.e., a (macroscopic) system will at all times be in one (and only one) of the states available to it] and \textit{non-invasive measurability}. Clearly quantum systems do not fulfill either of these assumptions and can thus violate Leggett-Garg inequalities. In contrast to Bell inequalities, where multiple spatially separated parties apply local operations, Leggett-Garg inequalities are obtained by probing the same system at multiple times.
We note that even if a Leggett-Garg inequality is violated, one can always argue that this happened because the measurement did influence the system, and therefore all subsequent measurements, in an unexpected (but in principle avoidable) way. This is known as the \textit{clumsiness} loophole. Since it is not possible to certify the non-invasiveness of a measurement, the clumsiness loophole can never be closed \cite{emary:2014}.

We will now show that the violation of a particular Leggett-Garg inequality directly implies the occurence of negative values in the corresponding KQPD. The considered inequality reads
\begin{equation}
\label{eq:leggettgarg}
K=C_{21}+C_{32}-C_{31}\leq 1,
\end{equation}
where the correlators are given by
\begin{equation}
\label{eq:lgcorr}
C_{ij} = \sum_{Q_i,Q_j=\pm 1}Q_iQ_jP_{ij}(Q_i,Q_j).
\end{equation}
Here $P_{ij}(Q_i,Q_j)$ is the probability of obtaining the outcome $Q_i$ at time $t_i$ and $Q_j$ at $t_j$ if the system is measured at these two times only. The measurement outcomes are restricted to $Q_i=\pm 1$. Violation of the inequality in Eq.~\eqref{eq:leggettgarg} implies that the correlators $C_{ij}$ can not be obtained from a positive probability distribution which describes all three measurements and is independent of the choice of measurements that are performed \cite{leggett:1985,emary:2014}, i.e. 
\begin{equation}
\label{eq:corrneq}
K>1\hspace{.5cm}\Rightarrow \hspace{.5cm}C_{ij}\neq \sum_{Q_1,Q_2,Q_3=\pm 1}Q_iQ_jP(Q_3,Q_2,Q_1),
\end{equation}
with $P(Q_3,Q_2,Q_1)\geq 0$.

For projective measurements of a dichotomic observable $\hat{Q}$, the correlator to be used in Eq.~\eqref{eq:leggettgarg} is given by \cite{fritz:2010}
\begin{equation}
\label{eq:corrproj}
C_{ij}={\rm Tr}\left\{\frac{1}{2}\left\{\hat{Q}(t_i),\hat{Q}(t_j)\right\}\hat{\rho}_0\right\}.
\end{equation}
We now introduce the KQPD for three subsequent measurements of $\hat{Q}$
\begin{equation}
\label{eq:kqpdlg}
\begin{aligned}
\mathcal{P}(Q_3,Q_2,Q_1)=&\frac{1}{(2\pi)^3}\int d\lambda_3d\lambda_2\lambda_1e^{i\lambda_3 Q_3+i\lambda_2 Q_2+i\lambda_1 Q_1}\\&\times{\rm Tr}\left\{e^{-i\frac{\lambda_3}{2}\hat{Q}(t_3)}e^{-i\frac{\lambda_2}{2}\hat{Q}(t_2)}e^{-i\frac{\lambda_1}{2}\hat{Q}(t_1)}\hat{\rho}_0e^{-i\frac{\lambda_1}{2}\hat{Q}(t_1)}e^{-i\frac{\lambda_2}{2}\hat{Q}(t_2)}e^{-i\frac{\lambda_3}{2}\hat{Q}(t_3)}\right\}.
\end{aligned}
\end{equation}
It is straightforward to show that the correlators in Eq.~\eqref{eq:corrproj} can directly be obtained from the KQPD
\begin{equation}
\label{eq:lgcorrkqpd}
C_{ij} = \int dQ_1dQ_2dQ_3\,Q_iQ_j\mathcal{P}(Q_3,Q_2,Q_1).
\end{equation}
If the KQPD was strictly positive, it would thus provide a positive probability distribution which describes all three measurements and is independent of the choice of measurements that are performed. The existence of such a probability distribution prevents the Leggett-Garg inequality to be violated \cite{emary:2014} (see also App.~\ref{app:lg}).
Inverting the argument, we conclude that a violation of the Leggett-Garg inequality in Eq.~\eqref{eq:leggettgarg} implies negative values for the KQPD in Eq.~\eqref{eq:kqpdlg} and thus non-classical behavior in the time-evolution of $\hat{Q}$.

{While we discussed projective measurements, the KQPD could also find applications when considering Leggett-Garg inequalities obtained from weak continuous measurements \cite{ruskov:2006,jordan:2006}.}

\section{Conclusions}
\label{sec:conclusions}
The KQPD provides a generalization of the Wigner function and the FCS to describe arbitrary observables within a quasi-probability formalism. This provides a unified framework for the description of fluctuations in dynamic quantum systems. Negative values of the KQPD can be directly linked to an interference between paths through Hilbert space which visit different eigenstates of the observables of interest. This requires coherent superpositions {of states that correspond to different measurement outcomes}, implying that negativity in the KQPD indicates non-classical behavior. As we have shown, this negativity is witnessed by anomalous weak values and violated Leggett-Garg inequalities (see also Refs.~\cite{williams:2008,groen:2013} for a connection between anomalous weak values and Leggett-Garg inequalities). The versatility of the KQPD allows to describe the problem at hand with a tailor-made quasi-probability distribution which focuses on the observables of interest.
When measuring these observables in von Neumann measurements, the outcome is determined by the KQPD, corrupted by measurement imprecision and back-action. 
However, the underlying KQPD can in principle be reconstructed tomographically by coupling the observables of interest to qubits and performing state tomography on the qubits. This has been discussed extensively for the FCS \cite{levitov:1996,clerk:2011,hofer:2016,dasenbrook:2016,lebedev:2016}.

{While we presented the point of view of the KQPD being a measurement-independent quantity that captures non-classicality of the system, a complementary viewpoint is provided by only considering von Neumann measurement setups. The appearance of negative values in the KQPD then reflects the presence of non-trivial correlations between the system and the detectors during the measurement.}

Here we only discussed a small number of applications for the KQPD. Other scenarios where the KQPD might be of interest include: measurement based quantum computing \cite{briegel:2009}; could the KQPD in this case capture the non-classical features that allow for outperforming a classical computer? Quantum-to-classical transition; how much noise is necessary for ensuring positivity of the KQPD and in which cases can the back-action in a measurement be neglected? Bell non-locality, entanglement, and steering; are these notions of non-classicality captured by a negative KQPD?
In addition to these open questions, we believe that there are many useful applications for the KQPD yet to be discovered.

\section*{Acknowledgments}
I acknowledge stimulating discussions with A. Clerk, N. Brunner, C. Flindt, and M. Perarnau-Llobet. I gratefully acknowledge the hospitality of McGill University where part of this work has been done. This work was funded by the Swiss National Science foundation.

\appendix

\section{Non-unitary time evolution}
\label{app:nonunitary}
In the main text, only unitary time evolution is considered. Generalizing the results to non-unitary time evolution is straightforward but leads to cumbersome expressions. Considering first $N$ operators observed at subsequent times, Eq.~\eqref{eq:momgenmultsub} has to be replaced by
\begin{equation}
\label{eq:momgenmultsubnonu}
\Lambda(\boldsymbol{\lambda})={\rm Tr}\left\{\mathcal{X}_N(\lambda_N)\mathcal{U}(\tau_N,\tau_{N-1})\cdots\mathcal{X}_2(\lambda_2)\mathcal{U}(\tau_2,\tau_1)\mathcal{X}_1(\lambda_1)\mathcal{U}(\tau_1,0)\hat{\rho}_0\right\},
\end{equation}
where the time evolution superoperator is defined for $\tau'\geq\tau$
\begin{equation}
\label{eq:timeevsup}
\hat{\rho}_{\tau'} =\mathcal{U}(\tau',\tau)\hat{\rho}_\tau,
\end{equation}
and 
\begin{equation}
\label{eq:xsup}
\mathcal{X}_j(\lambda_j)\hat{\rho}=e^{-i\frac{\lambda_j}{2}\hat{A}_j}\hat{\rho}e^{-i\frac{\lambda_j}{2}\hat{A}_j}.
\end{equation}
When considering operators over finite time-intervals, we divide the time interval into $N$ infinitesimal slices of width $\delta t$
 and insert a superoperator akin to $\mathcal{X}$ after each time slice. The completely general expression for the moment generating function for non-unitary time evolution then reads
\begin{equation}
\label{eq:momgennu}
\Lambda(\boldsymbol{\lambda})={\rm Tr}\left\{\mathcal{X}(t_N)\mathcal{U}(t_N, t_{N-1})\cdots\mathcal{X}(t_2)\mathcal{U}(t_2,t_1)\mathcal{X}(t_1)\mathcal{U}(t_1,0)\hat{\rho}_0\right\},
\end{equation}
where $t_j=j\delta t$ and
\begin{equation}
\label{eq:xsup2}
\mathcal{X}(t_j)\hat{\rho}=e^{-\frac{i}{2}\sum_l\delta tf_l(t_j)\lambda_l\hat{A}_l}\hat{\rho}e^{-\frac{i}{2}\sum_l\delta tf_l(t_j)\lambda_l\hat{A}_l}.
\end{equation}

\section{von Neumann measurements}
\label{app:vonneumann}
In this section, we calculate the probability distribution that describes a von Neumann measurement of subsequent instantaneous measurements. To this end, we first calculate the probability distribution for finding the detectors at the positions $r_j$, grouped in the vector $\boldsymbol{r}$, after interacting with the system
\begin{equation}
\label{eq:probadetpos}
\begin{aligned}
P(\boldsymbol{r})={\rm Tr}\bigg\{\langle \boldsymbol{r}|e^{-i\chi_N\hat{A}_N\hat{\pi}_N}\hat{U}(\tau_N,\tau_{N-1})&e^{-i\chi_{N-1}\hat{A}_{N-1}\hat{\pi}_{N-1}}\cdots e^{-i\chi_1\hat{A}_1\hat{\pi}_1}\hat{\rho}_{\tau_1}e^{i\chi_1\hat{A}_1\hat{\pi}_1}\cdots\\&\cdots e^{i\chi_{N-1}\hat{A}_{N-1}\hat{\pi}_{N-1}}\hat{U}(\tau_{N-1},\tau_{N}) e^{i\chi_N\hat{A}_N\hat{\pi}_N}|\boldsymbol{r}\rangle\bigg\}.
\end{aligned}
\end{equation}
Inserting identities resolved in the eigenstates of the $\hat{A}_j$ operators then yields
\begin{equation}
\label{eq:probadetpos2}
P(\boldsymbol{r})=\sum_{\boldsymbol{A}_L,\boldsymbol{A}_R}\left[\prod_l\langle r_l-\chi_lA_l^L|\hat{\rho}_l|r_l-\chi_lA_l^R\rangle\right]\delta_{A_N^L,A_N^R}\mathcal{A}(\boldsymbol{A}_L)\mathcal{A}^*(\boldsymbol{A}_R)\langle A^L_1|\hat{\rho}_{\tau_1}|A^R_1\rangle.
\end{equation}
Using the identity
\begin{equation}
\langle r_l-\chi_lA_l^L|\hat{\rho}_l|r_l-\chi_lA_l^R\rangle=\int \frac{d\gamma_l}{\chi_l}e^{-i\gamma_l(A_l^L-A_l^R)}\mathcal{W}_l(r_l-\chi_l[A_l^L+A_l^R]/2,\gamma_l/\chi_l),
\end{equation}
we find
\begin{equation}
\label{eq:probadetpos3}
P(\boldsymbol{r})=\int d\boldsymbol{A}'d\boldsymbol{\gamma}\left[\prod_l \mathcal{W}_l(r_l-\chi_lA_l',\gamma_l/\chi_l)/\chi_l\right]\mathcal{P}(\boldsymbol{A}';\boldsymbol{\gamma}).
\end{equation}
Finally, we recover Eq.~\eqref{eq:probasub} by noting that $P(\boldsymbol{r})d\boldsymbol{r}=P(\boldsymbol{A})d\boldsymbol{A}$ and consequently
\begin{equation}
P(\boldsymbol{A})=P(\boldsymbol{r})\big|_{\{r_j=\chi_jA_j\}}\prod_l\chi_l.
\end{equation}

\section{Leggett-Garg inequality violation implies negativity in the KQPD}
\label{app:lg}
We consider the KQPD given in Eq.~\eqref{eq:kqpdlg}. First, we convert the probability density distribution into a probability distribution ($\epsilon<1$)
\begin{equation}
\begin{aligned}
\tilde{\mathcal{P}}(Q_3,Q_2,Q_1)=&\sum_{\substack{Q_{1/2}=0,\pm 1\\Q_3=\pm 1}}\int_{Q_1-\epsilon}^{Q_1+\epsilon} dQ'_1\int_{Q_2-\epsilon}^{Q_2+\epsilon} dQ'_2\int_{Q_3-\epsilon}^{Q_3+\epsilon} dQ'_3\mathcal{P}(Q'_3,Q'_2,Q'_1)\\=&
\sum_{Q_{1/2}^{L/R}=\pm 1}\delta_{Q_1,(Q_1^L+Q_1^R)/2}\delta_{Q_2,(Q_2^L+Q_2^R)/2}\langle Q_3|\hat{U}(t_3,t_2)|Q_2^L\rangle\langle Q_2^L|\hat{U}(t_2,t_1)|Q_1^L\rangle\\&\hspace{3cm}\times\langle Q_1^L|\hat{\rho}_{t_1}|Q_1^R\rangle\langle Q_1^R|\hat{U}(t_1,t_2)|Q_2^R\rangle\langle Q_2^R|\hat{U}(t_2,t_3)|Q_3\rangle.
\end{aligned}
\end{equation}
Note that the KQPD can also be finite for $Q_{1/2}=0$ due to the terms where $Q^L_{1/2}\neq Q^R_{1/2}$.

We will now show that a purely positive KQPD implies that the Leggett-Garg inequality in Eq.~\eqref{eq:leggettgarg} can not be violated. First, note that
\begin{equation}
\label{eq:sumdiagkqpd}
\sum_{Q_1,Q_2, Q_3=\pm 1}\tilde{\mathcal{P}}(Q_3,Q_2,Q_1)=1.
\end{equation}
Together with the normalization of the KQPD, this implies
\begin{equation}
\label{eq:sumoffdiagkqpd}
\sum_{Q_1, Q_3=\pm 1}\tilde{\mathcal{P}}(Q_3,0,Q_1)+\sum_{Q_2, Q_3=\pm 1}\tilde{\mathcal{P}}(Q_3,Q_2,0)+\sum_{Q_3=\pm 1}\tilde{\mathcal{P}}(Q_3,0,0)=0.
\end{equation}
For a purely positive KQPD, all terms where at least one argument is zero therefore vanish. The correlators for projective measurements [cf.~Eq.~\eqref{eq:corrproj}] can then be written as
\begin{equation}
\label{eq:corrlgtil}
C_{ij} = \sum_{Q_1,Q_2,Q_3=\pm 1}Q_iQ_j\tilde{\mathcal{P}}(Q_3,Q_2,Q_1).
\end{equation}
For a positive KQPD, $\tilde{\mathcal{P}}$ thus provides a positive probability distribution which describes all three measurements and fulfills the assumptions of macroscopic realism and non-invasive measurements, preventing a violation of the Leggett-Garg inequality \cite{leggett:1985,emary:2014}.

%

\newpage
\bibliographystyle{quantum_ph}
\bibliography{biblio}

\end{document}